\documentclass{article}
\usepackage{amsmath}
\usepackage{hyperref}

\input{tcilatex}
\begin{document}

\title{The relation between Hawking radiation via tunnelling and the laws of
black hole thermodynamics}
\author{Yapeng Hu$^{\thanks{%
e-mail:huzhengzhong2050@163.com}}$, Jingyi Zhang$^{\thanks{%
e-mail:physicz@263.net}}$, Zheng Zhao$^{\thanks{%
e-mail:zhaoz43@hotmail.com}}$ \\
Department of Physics, Beijing Normal University, Beijing, 100875}
\maketitle

\begin{abstract}
In Parikh and Wilczek's original works, the laws of black hole
thermodynamics are not referred and it seems that there is no relation
between Hawking radiation via tunnelling and the laws of black hole
thermodynamics in their works. However, taking examples for the R-N black
hole and the Kerr black hole, we find that they are correlated and even
consistent if the tunnelling process is a reversible process.
\end{abstract}

\section{Introduction}

In 2000, Parikh and Wilczek reconsidered Hawking radiation. They argued that
the Hawking radiation is a tunnelling process and the barrier is created
just by the outgoing particle itself. In this way, They calculated the
emission rate from some static black holes and obtained the corrected
spectrum\cite{1,2,3,4}. Particularly, their result is consistent with the
underlying unitary theory and support to the information conservation.
Following this method, many other static and stationary rotating black holes
are also studied\cite{5,6,7,8,9,10,11,12,13,14}, and the results are the
same as that in Parikh and Wilczek's original works. In this paper, basing
on those works, and as a further study, we reinvestigate the Hawking
radiation via tunnelling. As we know, the laws of black hole thermodynamics
are not referred neither in Parikh and Wilczek's original works nor the
extended works based on them. And it seems that there is no relation between
Hawking radiation via tunnelling and the laws of black hole thermodynamics.
Is it true that there is no relation between them\cite{15}? In the following
discussion, we first take an example of the Reissner-Nordstrom black hole.
we reinvestigate the Hawking radiation of charged particles via tunnelling
from it\cite{6}, and rewrite the imaginary part of the action, which is
viewed from the laws of black hole thermodynamics. Then, in the same way we
reinvestigate the Hawking radiation of particles with angular momentum via
tunnelling from the Kerr black hole\cite{7}. Finally, we give a brief
conclusion and some discussion about our result.

\section{Taking the R-N black hole for example}

Recently, Parikh and Wilczek's original works have been extended to the R-N
black hole. As a typical static spherically symmetric solution of Einstein's
field equation, we first take it for example. According to Ref\cite{6}, the
imaginary part of the action for the classically forbidden trajectory is

\begin{eqnarray}
\func{Im}S &=&\func{Im}\{\int\nolimits_{r_{i}}^{r_{f}}[p_{r}-\frac{p_{A_{t}}%
\overset{\cdot }{A_{t}}}{\overset{\cdot }{r}}]dr\}  \notag \\
&=&-\func{Im}\{\int\nolimits_{r_{i}}^{r_{f}}\int_{(M,Q)}^{(M-\omega ,Q-q)}[%
\frac{2r\sqrt{2Mr-Q^{2}}}{r^{2}-2Mr+Q^{2}}dM-\frac{2\sqrt{2Mr-Q^{2}}Q}{%
r^{2}-2Mr+Q^{2}}dQ]dr\}  \notag \\
&=&-\pi \int_{(M,Q)}^{(M-\omega ,Q-q)}[\frac{(M+\sqrt{M^{2}-Q^{2}})^{2}}{%
\sqrt{M^{2}-Q^{2}}}dM-\frac{(M+\sqrt{M^{2}-Q^{2}})Q}{\sqrt{M^{2}-Q^{2}}}dQ] 
\notag \\
&=&-\frac{\pi }{2}\{[(M-\omega )+\sqrt{(M-\omega )^{2}-(Q-q)^{2}}]^{2}-[M+%
\sqrt{M^{2}-Q^{2}}]^{2}\}  \notag \\
&=&-\frac{1}{2}\Delta S_{BH}.  \label{1}
\end{eqnarray}%
where $A_{t}=\frac{Q}{r}$ is the first component of the 4-Dimensional
electromagnetic potential, and $p_{A_{t}}$ is the corresponding canonical
momentum conjugate.

Usually, investigating (1), we can easily obtain that the result is
consistent with the underlying unitary theory, so the information is
conserved. Our work is also to reinvestigate (1), which we will discuss in
the following.

As we know, for the R-N black hole, when a charged massive particle tunnels
across the event horizon, the mass and the charge of black hole will be
changed as a consequence. According to the first law of black hole
thermodynamics, the differential Bekenstein-Smarr equation of the R-N black
hole is\cite{16}

\begin{equation}
dM=\frac{\kappa }{8\pi }dA+VdQ\text{ }(J=0),  \label{2}
\end{equation}%
Furthermore, if the tunnelling process is considered as a reversible
process, according to the second law of black hole thermodynamics, (2) can
be rewritten as

\begin{equation}
dM=TdS+VdQ.  \label{3}
\end{equation}%
Equally, it can be rewritten as

\begin{equation}
dS=\frac{dM}{T}-\frac{VdQ}{T}.  \label{4}
\end{equation}

The temperature and the potential are respectively\cite{6}

\begin{equation}
T=\frac{\sqrt{M^{2}-Q^{2}}}{2\pi (M+\sqrt{M^{2}-Q^{2}})^{2}},V=\frac{Q}{M+%
\sqrt{M^{2}-Q^{2}}}.  \label{5}
\end{equation}%
Substituting (5) into (4), we can obtain

\begin{equation}
dS=\frac{2\pi (M+\sqrt{M^{2}-Q^{2}})^{2}}{\sqrt{M^{2}-Q^{2}}}dM-\frac{2\pi
(M+\sqrt{M^{2}-Q^{2}})Q}{\sqrt{M^{2}-Q^{2}}}dQ.  \label{6}
\end{equation}%
Thus, using (6), we can rewrite (1) as

\begin{eqnarray}
\func{Im}S &=&-\pi \int_{(M,Q)}^{(M-\omega ,Q-q)}[\frac{(M+\sqrt{M^{2}-Q^{2}}%
)^{2}}{\sqrt{M^{2}-Q^{2}}}dM-\frac{(M+\sqrt{M^{2}-Q^{2}})Q}{\sqrt{M^{2}-Q^{2}%
}}dQ]  \notag \\
&=&-\frac{1}{2}\int_{S_{i}}^{S_{f}}dS=-\frac{1}{2}\Delta S_{BH}.  \label{7}
\end{eqnarray}%
Which is also the same result as (1). The difference is that the result in
(7) implicates that Hawking radiation via tunnelling is correlated with the
laws of black hole thermodynamics.

\section{Taking the Kerr black hole for example}

Having taking the R-N black hole for example, instead, we study another
typical solution, a stationary axially symmetrical Kerr black hole. And for
the sake of simplicity, we only reinvestigate the case of a massless
particle with angular momentum tunnelling across the event horizon.
According to Ref\cite{7}, the imaginary part of the action is

\begin{eqnarray}
\func{Im}S &=&\func{Im}[\int_{r_{i}}^{r_{f}}p_{r}dr-\int_{\varphi
_{i}}^{\varphi _{f}}p_{\varphi }d\varphi ]  \notag \\
&=&\func{Im}[\int\nolimits_{r_{i}}^{r_{f}}\int_{M}^{M-\omega }\frac{\sqrt{%
(r^{2}+a^{2})^{2}-\Delta a^{2}\sin ^{2}\theta }}{\rho ^{2}-\sqrt{\rho
^{2}(\rho ^{2}-\Delta )}}drdM  \notag \\
&&-\int\nolimits_{r_{i}}^{r_{f}}\int_{M}^{M-\omega }\frac{\sqrt{%
(r^{2}+a^{2})^{2}-\Delta a^{2}\sin ^{2}\theta }}{\rho ^{2}-\sqrt{\rho
^{2}(\rho ^{2}-\Delta )}}a\Omega drdM]  \notag \\
&=&\int_{M}^{M-\omega }\frac{-2\pi (M^{2}+M\sqrt{M^{2}-a^{2}})}{\sqrt{%
M^{2}-a^{2}}}dM+\int_{M}^{M-\omega }\frac{\pi a^{2}}{\sqrt{M^{2}-a^{2}}}dM 
\notag \\
&=&\pi \lbrack M^{2}-(M-\omega )^{2}+M\sqrt{M^{2}-a^{2}}-(M-\omega )\sqrt{%
(M-\omega )^{2}-a^{2}}  \notag \\
&=&-\frac{1}{2}\Delta S_{BH}.  \label{8}
\end{eqnarray}

As the same discussion in section 2, Our next task is to view and rewrite
(8) from the laws of black hole thermodynamics. For the Kerr black hole, the
first law of black hole thermodynamics is\cite{16}

\begin{equation}
dM=\frac{\kappa }{8\pi }dA+\Omega dJ\text{ }(Q=0),  \label{9}
\end{equation}

And it will remain the same relationship if the tunnelling process is a
reversible process, we can obtain

\begin{equation}
dS=\frac{dM}{T}-\frac{\Omega dJ}{T}.  \label{10}
\end{equation}

The temperature, the angle velocity and the angular momentum of Kerr black
hole are respectively\cite{7}

\begin{equation}
T=\frac{\sqrt{M^{2}-a^{2}}}{4\pi (M^{2}+M\sqrt{M^{2}-a^{2}})},\Omega =\frac{a%
}{r_{+}^{2}+a^{2}}=\frac{a}{2(M^{2}+M\sqrt{M^{2}-a^{2}})},J=aM.  \label{11}
\end{equation}%
Thus, substituting (11) into (10), we get

\begin{equation}
dS=\frac{4\pi (M^{2}+M\sqrt{M^{2}-a^{2}})}{\sqrt{M^{2}-a^{2}}}dM-\frac{2\pi
a^{2}}{\sqrt{M^{2}-a^{2}}}dM.  \label{12}
\end{equation}%
Comparing (12) with (8), it is easy to find that we can also rewrite the
imaginary part of the action as follows

\begin{eqnarray}
\func{Im}S &=&\int_{M}^{M-\omega }\frac{-2\pi (M^{2}+M\sqrt{M^{2}-a^{2}})}{%
\sqrt{M^{2}-a^{2}}}dM+\int_{M}^{M-\omega }\frac{\pi a^{2}}{\sqrt{M^{2}-a^{2}}%
}dM  \notag \\
&=&-\frac{1}{2}\int_{M}^{M-\omega }[\frac{4\pi (M^{2}+M\sqrt{M^{2}-a^{2}})}{%
\sqrt{M^{2}-a^{2}}}dM-\frac{2\pi a^{2}}{\sqrt{M^{2}-a^{2}}}dM]  \notag \\
&=&-\frac{1}{2}\int_{S_{i}}^{S_{f}}dS=-\frac{1}{2}\Delta S_{BH}.  \label{13}
\end{eqnarray}%
Which shows that the Hawking radiation via tunnelling is also relative with
the laws of black hole thermodynamics for the stationary axial symmetry case.

\section{Conclusion and Discussion}

In section 2 and 3, we respectively take R-N black hole and Kerr black hole
for examples. And we obtain the result that the Hawking radiations via
tunnelling, both in the static spherically symmetric case and the stationary
axially symmetric case, are related to the laws of black hole
thermodynamics. In other words, Parikh and Wilczek's original works have
already made the assumption of the laws of black hole thermodynamics, in
detail, the first and the reversible second law of black hole
thermodynamics. More general speaking, Parikh and Wilczek's original works
are only suitable for the reversible process. However, in fact, because of
the negative heat capacity, an evaporating black hole is a highly unstable
system, and there is no stable thermal equilibrium between black hole and
the outside. Thus, the tunnelling process is usually an irreversible process
in principle. That is, it is a little early for Parikh and Wilczek to say
that the tunnelling process is consistent with the underlying unitary theory.

\section{Acknowledgements}

Thanks for Xuefei Gong's useful advice. This work is supported by the
National Natural Science Foundation of China under Grant No.10373003,
No.10475013 and the National Basic Research Program of China Grant
No.2003CB716300.

\end{document}